\documentclass[%
 reprint,
superscriptaddress,
 amsmath,amssymb,
 pra,
]{revtex4-2}

\usepackage{graphicx}
\usepackage{dcolumn}
\usepackage{bm}
\usepackage[caption=false]{subfig}
\usepackage{tikz}
    \usetikzlibrary[backgrounds]
    \usetikzlibrary[arrows.meta,bending,positioning]
    \usetikzlibrary{shapes.arrows}
\newcommand{\bb}{\mathbf}
\newcommand{\dd}{\,\mathrm{d}}

\usepackage[normalem]{ulem}

\begin{document}

\title{A local model for the optical energy and momentum transfer in dielectric media and the microscopic origin of Abraham’s force density}

\author{B. Anghinoni}
\email{brunoanghinoni@gmail.com}
\affiliation{Department of Physics, Universidade Estadual de Maring\'a, Maring\'a, PR 87020-900, Brazil}
\author{M. Partanen}
\affiliation{Department of Electronics and Nanoengineering, Aalto University, 00076 Aalto, Finland}
\author{N. G. C. Astrath}
\email{ngcastrath@uem.br}
\affiliation{Department of Physics, Universidade Estadual de Maring\'a, Maring\'a, PR 87020-900, Brazil}

\begin{abstract}
We report on the continuity equations for linear momentum and energy associated to a recently introduced electromagnetic formulation based on classical dipolar sources [Eur. Phys. J. Plus 138, 1034 (2023)]. When connected to the mass-polariton quasi-particle dynamics, these equations provide a consistent microscopic description of the local optical energy and momentum transfer inside dielectric media, called microscopic mass-polariton formulation. 
This procedure also unveils the true microscopic origin of the long-known Abraham optical force density as an interplay between induced dipoles and mechanical stresses generated within the material. 
\end{abstract}

\maketitle


\section{Introduction}\label{sec:intro}

Controlling opto-mechanical effects in matter is essential for numerous applications in physics, being especially important in optical manipulation techniques~\cite{Mansuripur2013b,Molloy2010,Gao2017,Ashkin1997,Shi2022,Li2020,Nieminen2007,Phillips1998}, photonics~\cite{Yang2022,Chin2020,Wiederhecker2009,Partanen2022b} and optofluidics~\cite{Psaltis2006,Monat2007,Garnier2003}. Currently, however, the electromagnetic forces acting inside matter when external fields are present do not have a definite description. This problem is closely related to the Abraham-Minkowski controversy~\cite{Brevik1979,Nelson1991,Pfeifer2007,Barnett2010,Kemp2011,Brevik2018c,Gordon1973,Milonni2010,Bethune-Waddell2015}, which treats the electromagnetic momentum transfer inside dielectrics and appeared more than a hundred years ago. Much of the difficulty associated with the Abraham-Minkowski controversy stems from the fact that probing flux and force densities and transfer rates of light in materials has been very challenging. The lack of successful experiments has led to interpretations that these densities and transfer rates would not in general be uniquely defined, but only integrated values would be physically meaningful \cite{Pfeifer2007}. However, advances in optics and photonics technologies are now starting to enable experimental setups, which make force densities of light measurable, at least indirectly \cite{Astrath2022,Astrath2023,Astrath2024}. From the theoretical side, different electromagnetic formulations exist in the literature to address this problem~\cite{Abraham,Minkowski,EL,Chu}, but no universal agreement has been reached yet.

Recently, we introduced the microscopic Ampère (MA) formulation~\cite{Astrath2022,Anghinoni2023}, which was developed for linear dielectric media under optical excitation. In this formulation, the medium is described as a continuum of classical point dipoles and the associated optical force density is
\begin{eqnarray}\label{eq:f3}
    \bb f_{\mathrm{MA}} = \frac{1}{2}\bm\nabla\left(\bb P \!\cdot\! \bb E \right)
 \!+\!\frac{1}{2}\bm\nabla\left(\bb M \!\cdot\! \bb B \right)
  \!-\!\frac{1}{2}|\bb E|^2\bm\nabla\varepsilon   \nonumber \\
  \!-\!\frac{1}{2}|\bb H|^2\bm\nabla\mu
  \!+\!\frac{n^2\!-\!1}{c^2}\frac{\partial}{\partial t}\left(\bb E \!\times\! \bb H\right), 
\end{eqnarray}
which is given in the laboratory frame and is valid for linear, isotropic, lossless media with no dispersion, assuming the atoms' velocities are much smaller than the velocity of light, $c$. In the last equation, $\bb E$ is the electric field, $\bb P$ is the polarization, $\bb H$ is the magnetic field, $\bb M$ is the magnetization, $\bb B = \mu_0(\bb M+\bb H)$ is the magnetic induction field, $\varepsilon$ and $\mu$ are the medium's permittivity and permeability, respectively, $t$ is time, and $n=\sqrt{\varepsilon \mu / \varepsilon_0 \mu_0}$ is the medium's refractive index, with $\varepsilon_0$ and $\mu_0$ denoting the vacuum's permittivity and permeability, respectively.

The MA force density in Eq.~(\ref{eq:f3}) is capable of describing the vast majority of experiments up to date involving optical forces and radiation pressure in dielectric media as the main physical effects, as discussed in detail in Ref.~\cite{Anghinoni2023}. The formalism has the advantage of being derived from a well-established microscopic model -- the classical dipolar sources. On the other hand, it was shown in Ref.~\cite{MP1} and subsequent works~\cite{MP2,MP4,MP5,MP6} that, in order to fulfill the covariance requirements from special relativity, there must be a coupled state of field and matter propagating through the dielectric -- the so-called mass-polariton (MP) quasi-particle, composed of the electromagnetic wave plus a mass density wave (MDW) generated by the disturbances in the atomic positions due to the optical force. More specifically, the MP formulation was built using the Abraham force density~\cite{Abraham,Anghinoni2022}, which has been long adopted to describe external electromagnetic forces acting in dielectrics~\cite{Milonni2010,LL}. Despite its relatively wide range of application, Abraham's force density is known to not contemplate the electro- and magnetostriction effects~\cite{Brevik1979,Partanen2023}, which are quadratic effects on the fields tending to compress the material towards the region of higher field intensity and correspond to the first and second terms in Eq.~(\ref{eq:f3}), respectively. This fact renders Abraham's force density unable to model experiments such as the ones reported in Refs.~\cite{Astrath2022,Astrath2023,Hakim1962,Zimmerli1999,Astrath2024} without further introducing (often phenomenologically) extra contributions to the related continuity equations. In truth, to the best of our knowledge, Eq.~(\ref{eq:f3}) is the only one in the literature in accordance with the measurements of the optical electrostriction force density reported in Refs.~\cite{Astrath2022,Astrath2023,Astrath2024} while also being compatible with the classical model of tiny current loops (also known as Ampèrian model) for the microscopic magnetization mechanism in matter~\cite{Jackson1977,Anghinoni2023}. Another very important fact is that the few experiments that were truly able to measure the photon momentum in matter showed a linear dependence on the medium's refractive index~\cite{Jones1954,Jones1978,Campbell2005,Gibson1980,Strait2019}, exactly as predicted by the coupled mass-polariton state~\cite{MP1}.

In the context just presented, it is natural to consider the possibility of properly incorporating the MA force density into the MP formulation -- i.e., to build the continuity equations for energy and momentum for the field+matter coupled system by using the MP quasi-particle while retaining the force density from Eq.~(\ref{eq:f3}). 
This unification of the models would provide a single consistent and non-phenomenological model for the local energy and momentum transfer inside dielectric media. This will be developed in the present work. Our model also reveals 
the microscopic origin of the Abraham optical force density in dielectrics as a byproduct. The results cover an extensive variety of optomechanical applications and further elucidate the centenary Abraham-Minkowski controversy on the photonic momentum inside matter. 

This work is organized as follows. In Sec.~\ref{sec:MA_SET}, we build the total stress-energy-momentum tensor of the coupled field and matter system. In Sec.~\ref{sec:cont}, we derive the closed continuity equations for the system by using the MP quasi-particle. In Sec.~\ref{sec:disc}, the general validity of the theory is discussed in detail and compared to the existing literature on the topic. In particular, the long-known Abraham force density is shown to have originated from an interplay between induced dipoles and mechanical stresses. At last, conclusions and future perspectives are drawn in Sec.~\ref{sec:conc}.

\section{Stress-energy-momentum tensor}\label{sec:MA_SET}

In flat space-time, the principles of conservation of energy and linear momentum can be described simultaneously by the covariant four-continuity equation \cite{LL_fields} $\partial_\nu \mathcal{T}^{\mu\nu} = - f^{\mu}$, where $\partial_{\nu}=(c^{-1}\partial_t,\partial_x,\partial_y,\partial_z)$, the indices $\mu,\nu=0,1,2,3$ denote the space-time index and $\mathcal{T}$ is the stress-energy-momentum (SEM) tensor, generically given as
\begin{equation}
\mathcal{T} =
\begin{pmatrix}
 W & \bb S/c \\
 c \bb g & \overleftrightarrow{\bb T} \\
\end{pmatrix}
,
\end{equation}
where $\overleftrightarrow{\mathbf{T}}$ is the stress tensor, $\mathbf{g}$ is the momentum density, $\bb{S}$ is the energy flux and $W$ is the energy density. Throughout this work, the Minkowski metric $\eta^{\mu\nu}$ in $(-,+,+,+)$ signature is implied. The four-force, on its turn, is $f^{\mu} = (\phi/c, \bb f)$, where $\phi$ is the power density and $\bb f$ is the force density. We must always have $\partial_\nu \mathcal{T}^{\mu\nu} = 0$ for a closed system.

\subsection{The field SEM tensor}\label{sec:MP+MA0}

The Lorentz force density law is given as $\bb f = \rho \bb E +\bb J \times \bb B$, where $\rho$ and $\bb J$ are the charge and current densities, respectively. The MA force density in Eq.~(\ref{eq:f3}) was derived by introducing into the Lorentz force density the four-current density associated to a single medium element (for example, an atom or a molecule) endowed with electric and magnetic dipole moments, namely~\cite{Zangwill2013,Griffiths2015}
\begin{equation}\label{eq:4J} 
    J^{\nu}_{\mathrm{MA}} = (-c(\bb p\cdot\bm\nabla)\delta^3(\bb r),\dot{\bb p}\delta^3(\bb r)-(\bb m\times\bm\nabla)\delta^3(\bb r)),
\end{equation}
where $\bb p$ is the electric dipole moment, $\bb m$ is the magnetic dipole moment, $\delta^3(\bb r)$ the three-dimensional Dirac delta distribution, the over-dot denotes the total time derivative and $\bb r$ is the position vector. As the force density from Eq.~(\ref{eq:f3}), the four-current shown in Eq.~(\ref{eq:4J}) is valid for the laboratory frame and non-relativistic velocities. Alternatively, it is also possible to obtain the optical force density from Eq.~(\ref{eq:f3}) from a full Lagrangian approach~\cite{Astrath2024} which does not need to invoke Eq.~(\ref{eq:4J}) explicitly. 

The procedure described above of defining an appropriate model for the electromagnetic sources in matter and then inserting them into the Lorentz force law is quite standard. For example, the conventional Ampère formulation for macroscopic fields~\cite{Anghinoni2022} is also derived in this fashion, where the effective, bound four-current $J^{\nu}_{\mathrm{b}}=(-c\bm\nabla\cdot \bb P, \bm\nabla\times\bb M +\partial_t \bb P)$ is employed. In fact, we notice that the widely known derivation of Maxwell's stress tensor starting from the Lorentz force density can be applied to any four-current configuration. This happens because, excluding self-field effects, it is always possible to substitute the electromagnetic sources by the corresponding electromagnetic fields in the Lorentz equation through Gauss' and Ampère-Maxwell's law, generating $\bb f = \varepsilon_0(\bm \nabla \cdot \bb E) \bb E + \mu_0^{-1}(\bm \nabla \times \bb B) \times \bb B$ regardless of the microscopic sources' details. We therefore conclude that Maxwell's SEM tensor, given by~\cite{LL_fields}
\begin{align}\label{eq:T_MA}
 &\mathcal{T}_{\mathrm{Maxwell}}\nonumber\\
&=\begin{pmatrix}
  W_\mathrm{Maxwell}  & \varepsilon_0 c(\bb E \times \bb B)^{\mathrm{T}} \\
\varepsilon_0 c\bb E \times \bb B     & \quad W_\mathrm{Maxwell}\overleftrightarrow {\bb I} -\varepsilon_0\bb E\!\otimes\!\bb E-\mu_0^{-1}\bb B\!\otimes\!\bb B
\end{pmatrix},
\end{align}
must be valid for any four-current configuration -- even though it does not contain the source terms explicitly. Here ``T'' denotes matrix transposition, ``$\otimes$'' denotes outer product, $\overleftrightarrow {\bb I}$ is the unit dyadic and $W_\mathrm{Maxwell}=(\varepsilon_0 |\bb  E|^2+\mu_0^{-1}|\bb B|^2)/2$. 


The SEM tensor in Eq.~(\ref{eq:T_MA}) corresponds to the field part of our system. Together with the SEM tensor of the material (which will be shown in the next section), they form a closed system whose four-divergence must be zero. Notice that in this SEM description there are no explicit interaction terms (as there would generally exist in Lagrangian descriptions) -- rather, the interaction term is related to the fact that the four-divergences of the two SEM tensor terms are nonzero four-force densities with opposite signs -- whence the four-divergence of the total SEM tensor is zero \cite{MP4,MP5,MP6}. Thus, interactions modify the values of the fields, but they are not explicitly seen in the SEM tensors. 
In our case, the four-divergence of Maxwell's SEM generates the force density $\bb f_{\mathrm{MA}}$ given in Eq.~(\ref{eq:f3}) and the power density $\phi_{\mathrm{MA}}= \bb E \cdot \bb J_{\mathrm{MA}}$ (see Appendix~\ref{sec:app_MA}), with $\bb J_{\mathrm{MA}}$ given by the spatial part of the four-current in Eq.~(\ref{eq:4J}). 
Contrary to Maxwell's SEM tensor, these interactions depend explicitly on the specific model adopted to describe the electromagnetic sources within the system. To calculate $\phi_{\mathrm{MA}}$, we start by writing the total power as 
\begin{equation}\label{eq:phi}
    \Phi_{\mathrm{MA}} = \int \bb E \cdot \left[\dot{\bb p}\delta^3(\bb r) - \bb m \times \bm \nabla \delta^3(\bb r)\right] \dd ^3 \bb r.
\end{equation}
%
%
%
%
Equation~(\ref{eq:phi}) corresponds to the contribution of a single medium element to the total power. When a sum of similar terms from medium elements at different positions is integrated over a small volume inside the dielectric (but still microscopically large, i.e., containing a large number of dipoles), and the dipoles' velocities are assumed much smaller than $c$, the calculation yields the power density as (see Appendix~\ref{app:phi} for the calculation)
\begin{equation}\label{eq:phi_ma}
    \phi_{\mathrm{MA}} = \frac{1}{2}\frac{\partial}{\partial t}\left({\bb P} \cdot \bb E -\bb M \cdot  \bb B\right),
\end{equation}
where $\phi_{\mathrm{MA}}$, $\bb P$ and $\bb M$ are given by $\Phi_{\mathrm{MA}}$, $\bb p$ and $\bb m$ divided by the integration volume and multiplied by the number of dipoles in this volume, respectively. 

Assuming no external fields are initially present, if we integrate $\phi_{\mathrm{MA}}$ in time we obtain the energy density $W_{\mathrm{int}}=({\bb P} \cdot \bb E -\bb M \cdot  \bb B)/2$, which represents the (negative of the) total work per unit volume that must be supplied to assemble the fields+dipoles configuration and to establish the dipoles within the fields. This is different from the well-known potential energy density of linear, stationary dipoles in external fields, $-({\bb P} \cdot \bb E +\bb M \cdot  \bb B)/2$, because to assemble the configuration extra work must be supplied to overcome the effects of the induced electric current, according to Lenz law. For a detailed discussion on this subtlety see, for example, Refs.~\cite{Jackson1999,Zangwill2013}.

With Eq.~(\ref{eq:phi_ma}), the four-continuity equation associated to the MA formulation, $\partial_{\nu}\mathcal{T}_{\mathrm{Maxwell}}^{\mu\nu} = -f_{\mathrm{MA}}^{\mu}$, is then completely defined. This equation contains the information of the electromagnetic fields dynamics and how they interact with the material -- therefore, by finding the appropriate SEM tensor for the material we are able to delineate our closed system. This tensor for the material will be discussed in the next section by using the mass-polariton quasi-particle theory~\cite{MP1}.

\subsection{The material and MDW SEM tensors}\label{sec:MP+MA}

The original MP formulation~\cite{MP1} adopts, in the laboratory frame, Abraham's force density $\bb f_{\mathrm{Ab}}=-|\bb E|^2\bm\nabla\varepsilon/2-|\bb H|^2\bm\nabla\mu/2+c^{-2}(n^2-1)\partial_t (\bb E \times \bb H)$ and the SEM tensor relation $\mathcal{T}_{\mathrm{MP}}=\mathcal{T}_{\mathrm{Ab}}+\mathcal{T}_{\mathrm{MDW}}$. In contrast to Maxwell's SEM tensor in Eq.~\eqref{eq:T_MA}, the original MP formulation used Abraham's SEM tensor, given by~\cite{Abraham}
\begin{equation}\label{T_Ab}
    \mathcal{T}_{\mathrm{Ab}}=
    \begin{pmatrix}
  W_\mathrm{Ab}  & (\bb E \times \bb H)^{\mathrm{T}}/c \\
\bb E \times \bb H/c     & \quad W_\mathrm{Ab}\overleftrightarrow {\bb I} -\bb D\otimes\bb E-\bb B\otimes\bb H
\end{pmatrix},
\end{equation}
where $W_\mathrm{Ab}=(\bb D \cdot \bb E+\bb B\cdot \bb H)/2$, and $\bb D = \varepsilon_0\bb E+\bb P$ is the electric displacement field. The MDW SEM tensor $\mathcal{T}_{\mathrm{MDW}}$ is discussed later in this section.

Before description of the MDW, we discuss the SEM tensor of the material, $\mathcal T_{\mathrm{mat}}$, which is given as a sum of a mass component and a mechanical component, namely $\mathcal T_{\mathrm{mat}}=\mathcal{T}_{\mathrm{mass}}+\mathcal{T}_{\mathrm{mech}}$. The SEM tensor of the material mass is given by~\cite{LL_fields}
{\begin{align}\label{eq:Tmass}
    \mathcal{T}_{\mathrm{mass}}^{\mu\nu} &=
    (\rho_\mathrm{a}/\gamma_{\bb v_\mathrm{a}}^2)U_\mathrm{a}^{\mu}U_\mathrm{a}^{\nu} =
\begin{pmatrix}
\rho_{\mathrm{a}}c^2     & \rho_{\mathrm{a}}\bb{v}_{\mathrm{a}}^{
\mathrm{T}}c \\
 \rho_{\mathrm{a}}\bb{v}_{\mathrm{a}}c     & \rho_{\mathrm{a}}\bb{v}_{\mathrm{a}}\otimes\bb{v}_{\mathrm{a}}
\end{pmatrix},
\end{align}
where $\rho_\mathrm{a}=\gamma_{\bb v_\mathrm{a}}^2\rho_\mathrm{ai0}+W_\mathrm{int}/c^2$ is the mass density of the material, in which $\rho_\mathrm{ai0}$ is the instantaneous rest mass density, $\gamma_{\bb v_\mathrm{a}} = (1-|\bb v_\mathrm{a}|^2/c^2)^{-1/2}$ is the Lorentz factor corresponding to the atomic velocity $\bb v_{\mathrm{a}}$, and $U_\mathrm{a}^{\mu}=\gamma_{\bb v_\mathrm{a}} (c,\bb v_\mathrm{a})$ is the four-velocity of the material. Under dynamical interactions, which locally vary the positions of atoms, $\rho_\mathrm{ai0}$ can differ from the rest mass density of the material at equilibrium with no interactions, which we denote by $\rho_\mathrm{a0}$. As discussed in Ref.~\cite{Anghinoni2022}, the relation of these two quantities is given by $\rho_\mathrm{ai0}=\rho_\mathrm{a0}/(1+\bm\nabla\cdot\mathbf{r}_\mathrm{a})$, where $\mathbf{r}_\mathrm{a}$ is the atomic displacement field averaged over a small volume, which however contains many atoms. In the original MP formulation \cite{MP1}, the energy of induced dipoles was considered as a part of Abraham's energy density $W_\mathrm{Ab}$, and consequently $W_\mathrm{int}$ was set to zero in the expression of $\rho_\mathrm{a}$ above. In the present theory, $W_\mathrm{int}$ is associated with establishing induced dipoles within the time-dependent electromagnetic field by the power density in Eq.~\eqref{eq:phi_ma} as discussed above. Thus, $W_\mathrm{int}$ is obtained by the time-integral of the power density of Eq.~\eqref{eq:phi_ma}, and it is equal to the difference of the Abraham energy density $W_\mathrm{Ab}$ of the original MP formulation and the energy density $W_\mathrm{Maxwell}$ of the present formulation as $W_\mathrm{int}=\int_{-\infty}^t\phi_\mathrm{MA}dt'=W_\mathrm{Ab}-W_\mathrm{Maxwell}$.

The explicit form of the mechanical SEM tensor, $\mathcal{T}_\mathrm{mech}$, depends on the specific deformation characteristics of the material -- e.g., a fluid or an elastic solid. For example, for a perfect fluid we have $\mathcal{T}_\mathrm{mech}^{\mu\nu} = (\mathcal P/c^2)U_\mathrm{a}^{\mu}U_\mathrm{a}^{\nu}+\mathcal{P}\eta^{\mu\nu}$ \cite{LL_fields,Luscombe}, where $\mathcal{P}$ denotes the local mechanical pressure. 

The MDW SEM tensor is given for non-relativistic atomic velocities in the laboratory frame as a difference of the SEM tensor of the material mass in Eq.~\eqref{eq:Tmass} disturbed from equilibrium by the electromagnetic force and power densities and the same formula with $\bb v_\mathrm{a}=0$ and $W_\mathrm{int}=0$, denoted by $\mathcal{T}_\mathrm{mass,0}^{\mu\nu}$, as~\cite{MP1,MP4,MP5,MP6}
\begin{align} \label{eq:TMDW}
    \mathcal{T}_{\mathrm{MDW}} &=\mathcal{T}_\mathrm{mass}^{\mu\nu}-\mathcal{T}_\mathrm{mass,0}^{\mu\nu}\nonumber\\
&=
\begin{pmatrix}
\rho_{\mathrm{MDW}}c^2     & \rho_{\mathrm{a}}\bb{v}_{\mathrm{a}}^{
\mathrm{T}}c \\
 \rho_{\mathrm{a}}\bb{v}_{\mathrm{a}}c     & \rho_{\mathrm{a}}\bb{v}_{\mathrm{a}}\otimes\bb{v}_{\mathrm{a}}
\end{pmatrix}
.
\end{align}
In this equation, the mass density of the atomic MDW is defined as $\rho_{\mathrm{MDW}} = \rho_{\mathrm{a}} - \rho_{\mathrm{a0}}=\gamma_{\bb v_\mathrm{a}}^2\rho_\mathrm{ai0} - \rho_{\mathrm{a0}}+W_\mathrm{int}/c^2$. The kinetic energy of atoms, resulting from optical forces, is extremely small due to its quadratic dependence on the small atomic velocity in the assumed non-relativistic regime and hence can be neglected~\cite{MP6}. Thus, we can set $\gamma_{\bb v_\mathrm{a}}^2=1$ in the equations above. Using the relation $\rho_\mathrm{ai0}=\rho_\mathrm{a0}/(1+\bm\nabla\cdot\mathbf{r}_\mathrm{a})$ in the limit of small $\bm\nabla\cdot\mathbf{r}_\mathrm{a}$, we then obtain that the MDW mass density in the laboratory frame can be approximated as
\begin{equation}\label{eq:rhoMDW}
    \rho_\mathrm{MDW}=\rho_{\mathrm{a}} - \rho_{\mathrm{a0}}
    =-(\bm\nabla\cdot\mathbf{r}_\mathrm{a})\rho_\mathrm{a0}+W_\mathrm{int}/c^2.
\end{equation}
The first term is related to atomic displacements, and the second term, which was absent in the original MP formulation, is the mass equivalent of the interaction energy associated with the creation of induced dipoles by the power density of Eq.~\eqref{eq:phi_ma}.


\subsection{Incorporating the mass-polariton dynamics}\label{sec:MP+MA2}

The MA force density is the sum of Abraham's force density and the electro- and magnetostriction effects, i.e., 
\begin{equation}\label{eq:fma+fab}
 \bb f_{\mathrm{MA}}=\bb f_{\mathrm{Ab}}+\bm\nabla(\bb P\cdot\bb E)/2+\bm\nabla(\bb M\cdot\bb B)/2.    
\end{equation}
In terms of stress tensors and momentum densities, the last equation should be equivalent to
\begin{eqnarray}\label{eq:fma_ab}   \overleftrightarrow{\bm\nabla}\cdot\left(\overleftrightarrow{\bb T}_{\mathrm{Ab}}-\overleftrightarrow{\bb T}_{\mathrm{Maxwell}}\right)+\frac{\partial }{\partial t}\left(\bb g_{\mathrm{Ab}}-\bb g_{\mathrm{Maxwell}}\right)=\nonumber \\
   \frac{1}{2}  \overleftrightarrow{\bm\nabla}\cdot\left[(\bb P \cdot \bb E)\overleftrightarrow{\bb I} + (\bb M \cdot \bb B)\overleftrightarrow{\bb I}\right],
\end{eqnarray}
where $\overleftrightarrow{\bb T}_{\mathrm{Ab}}=W_\mathrm{Ab}\overleftrightarrow {\bb I} -\bb D\!\otimes\!\bb E-\bb B\!\otimes\!\bb H$, $\overleftrightarrow{\bb T}_{\mathrm{Maxwell}}=W_\mathrm{Maxwell}\overleftrightarrow {\bb I} -\varepsilon_0\bb E\!\otimes\!\bb E-\mu_0^{-1}\bb B\!\otimes\!\bb B$, $\bb g_{\mathrm{Ab}} = \bb E \times \bb H/c^2 $ and $\bb g_{\mathrm{Maxwell}} = \varepsilon_0 \bb E \times \bb B $. It can be verified that Eq.~(\ref{eq:fma_ab}) does not hold as the magnetic part is unable to satisfy the relation. The subtle issue to be noticed here is that $\bb f_{\mathrm{MA}}$ and $\bb f_{\mathrm{Ab}}$ were actually constructed for different physical systems. The force density $\bb f_{\mathrm{MA}}$ is built from point dipoles and describes the electromagnetic interactions locally, whereby ``locally'' we mean in a small volume inside the dielectric, which contains many dipoles (i.e., is microscopically large) but is still smaller than the total volume of the material. This feature is very important as it allows the MA formulation to describe the spatio-temporal dependence of the optical forces in dielectric media. The force density $\bb f_{\mathrm{Ab}}$, on its turn, is obtained from the equation $\bb f_{\mathrm{Ab}}= - \overleftrightarrow{\bm \nabla} \cdot \overleftrightarrow{\bb {T}}_{\mathrm{Ab}}-\partial_t\bb g_{\mathrm{Ab}}$ by employing Gauss' and Ampère-Maxwell's laws in their macroscopic forms~\cite{Milonni2010}, i.e., $\bm \nabla \cdot \bb D = \rho_\mathrm f$ and $\bm \nabla \times \bb H = \bb J_\mathrm f + \partial_t \bb D$, respectively (where $\rho_\mathrm f$ and $\bb J_\mathrm f$ are the free electric charge and current density, respectively). In these macroscopic forms the effective bound four-current density $J^{\nu}_{\mathrm{b}}=(-c\bm \nabla \cdot \bb P, \bm \nabla\times\bb M+\partial_t \bb P)$ is automatically assumed -- therefore, formulations based on this electromagnetic source model can not be applied to describe local optical forces, being appropriate only if one is interested in the overall center-of-mass motion of the material~\cite{Brevik1979,Anghinoni2022}. 
Thus, in local descriptions of optical forces in matter, Abraham's force density can not be applied. 
In fact, the same problem also happens to the Einstein-Laub formulation~\cite{EL}, which, although initially modeled through point dipoles as well, also employs the macroscopic Maxwell's equations in the derivation of its continuity equations~\cite{Brevik1979, Anghinoni2022}. 
This is a very subtle and important fact for the Abraham-Minkowski debate, which seems to have been overlooked in the literature.
Again, we stress that the only way to treat the electromagnetic sources generically, be them bound and/or free, is by means of Maxwell's SEM tensor, as discussed at the beginning of this section. 

In the scenario discussed, we propose an alternative form of the MP tensor, called microscopic mass-polariton (MMP) tensor and given as 
\begin{equation}\label{eq:T_MMP}   \mathcal{T}_{\mathrm{MMP}}=\mathcal{T}_{\mathrm{Maxwell}}+\mathcal{T}_{\mathrm{MDW}}, 
\end{equation}
where the SEM tensors $\mathcal{T}_{\mathrm{Maxwell}}$ and $\mathcal{T}_{\mathrm{MDW}}$ are given in Eqs.~(\ref{eq:T_MA}) and (\ref{eq:TMDW}), respectively. As stated earlier, $\mathcal{T}_{\mathrm{MDW}}$ encompasses the out-of-equilibrium rest mass and kinetic energies of the system. The remaining mechanical effects are described by the tensor $\mathcal{T}_{\mathrm{mech}}$, which was explicitly written in the beginning of this section for a perfect fluid as an example.

The total SEM tensor of the closed system in our model is therefore given by $\mathcal{T}_{\mathrm{tot}} = \mathcal{T}_{\mathrm{Maxwell}}+\mathcal{T}_{\mathrm{mat}}$, or
\begin{align}\label{eq:T_tot} \mathcal{T}_{\mathrm{tot}}&=\mathcal{T}_{\mathrm{Maxwell}}+\mathcal{T}_{\mathrm{mass}}+\mathcal{T}_{\mathrm{mech}}\nonumber\\
&=\mathcal{T}_{\mathrm{MMP}}+\mathcal{T}_{\mathrm{mass,0}}+\mathcal{T}_{\mathrm{mech}}. 
\end{align} 
In mechanical equilibrium in the presence of static electric and magnetic fields, $\mathcal{T}_{\mathrm{mech}}$ is counterbalanced within the dielectric by the electro- and magnetostriction force densities, as will be discussed in the following section. For time-varying electromagnetic fields, on the other hand, the striction force densities act as sources in the acoustic pressure wave equation~\cite{Tam1986}, being therefore associated to the transient acoustic relaxation of the medium.

\section{Continuity equations}\label{sec:cont}
As the electromagnetic field and the material in our case form a closed system, we must have $\partial_{\nu}\mathcal{T}_{\mathrm{tot}}^{\mu\nu}=0$. This equation can be conveniently split into its temporal ($\mu =0$) and spatial ($\mu=1,2,3$) coordinates, providing explicitly the energy and momentum continuity equations for our theory, respectively, which will be shown next.

\subsection{Energy continuity equation}\label{sec:en}

Using the explicit forms of the SEM tensors and $\partial_\nu=(c^{-1}\partial_t,\partial_x,\partial_y,\partial_z)$, we obtain the energy continuity equation in the laboratory frame as
\begin{eqnarray}\label{eq:MPMA_en}    
\partial_{\nu}\mathcal T^{0\nu}_{\mathrm {tot}}=\frac{1}{c}\frac{\partial}{\partial t}\left(\frac{1}{2}\varepsilon_0 |\bb  E|^2 + \frac{1}{2\mu_0}|\bb B|^2+\rho_{\mathrm{a}}c^2\right)\nonumber \\
    +\frac{1}{c}\bm\nabla\cdot\left(\bb E \times \bb H+\rho_{\mathrm{a}}\bb v_{\mathrm{a}}c^2\right)=0.
\end{eqnarray}
Here, the hidden energy contribution~\cite{Shockley1967} $\bb S_{\mathrm{h}}=\bb M \times \bb E$ (also known as Aharonov-Casher interaction~\cite{Aharonov1984}) has been added to the electromagnetic contribution of the energy flux as $\bb E \times \bb H = \mu_0^{-1} \bb E\times\bb B + \bb M \times \bb E=\bb S_{\mathrm{Maxwell}}+\bb S_{\mathrm{h}}$. This is necessary because, as discussed in detail in Ref.~\cite{Anghinoni2023}, the derivations of the MA force density are non-relativistic and therefore need the inclusion of the hidden momentum and energy contributions in an \emph{ad hoc} manner, when writing equations for the laboratory frame, to keep the correct relativistic transformation properties~\cite{Hnizdo1997} -- had we employed a relativistic derivation from the start, these terms would arise naturally when the center of mass-energy of the dipole system is promoted to a dynamical variable, as shown in Ref.~\cite{Horsley2006} 
(see Appendix~\ref{sec:HM} for more details).
By adding these contributions to the four-continuity equation in the laboratory inertial reference frame it is implied we are keeping contributions only to order $|\bb v_{\mathrm{a}}|/c$ -- therefore, the relatively small atomic kinetic energy is neglected, as stated earlier. 
As the motion of the atoms is accelerated due to the external forces, their proper frames are not inertial reference frames -- the velocity $\bb v_{\mathrm{a}}$ is the relative velocity between a local inertial reference frame, moving along with the atoms, and the laboratory frame~\cite{MP5}.

Notice that Eq.~(\ref{eq:MPMA_en}) has a distinct form from the original MP formulation, which in this case would be $\partial_t (W_{\mathrm{Ab}}+\rho_\mathrm a c^2)+\bm\nabla\cdot(\bb E \times \bb H +\rho_\mathrm a \bb v_\mathrm a c^2)=0$~\cite{MP1}. The difference stems from the field energy density $W_{\mathrm{Maxwell}} = (\varepsilon_0 |\bb  E|^2+\mu_0^{-1}|\bb B|^2)/2$ appearing in the time derivative term instead of $W_{\mathrm{Ab}}$ (see Eq.~(\ref{T_Ab})). To elucidate this discrepancy, we first note that, from the Sec.~\ref{sec:MA_SET}, we have $\partial_{\nu}\mathcal T^{0\nu}_{\mathrm{Maxwell}}=-\phi_{\mathrm{MA}}/c$. Using the last form of Eq.~\eqref{eq:T_tot}, we then have 
\begin{eqnarray}\label{eq:en2}
  \partial_{\nu}\mathcal T^{0\nu}_{\mathrm{tot}}=-\phi_{\mathrm{MA}}+\partial_t(\rho_{\mathrm{a}}c^2)+\bm\nabla\cdot(\rho_{\mathrm{a}}\bb v_{\mathrm{a}}c^2)=0,  
\end{eqnarray}
which is an alternative form of Eq~(\ref{eq:MPMA_en}). As previously stated, in our system the power density $\phi_{\mathrm{MA}}$ is related to the assembly and establishment of the induced dipoles within the fields, with an associated energy density equal to $W_{\mathrm{int}} =(\bb P\cdot\bb E-\bb M\cdot\bb B)/2$. As discussed in Sec.~\ref{sec:MP+MA}, this energy density becomes implicitly contained in $\rho_{\mathrm{a}}c^2$. From the point of view of Eq.~(\ref{eq:en2}), this fact can be seen as follows. First, we approximate $\rho_{\mathrm{a}}\bb v_{\mathrm{a}}\approx \rho_{\mathrm{a0}}\bb v_{\mathrm{a}}=\rho_{\mathrm{a0}}\dd\bb r_\mathrm{a}/\dd t\approx\rho_{\mathrm{a0}}\partial_t\bb r_\mathrm{a}$. Thus, Eq.~(\ref{eq:en2}) becomes $\partial_t[\rho_\mathrm{a}c^2+(\bm\nabla\cdot\bb r_\mathrm{a})\rho_\mathrm{a0}c^2]=\phi_{\mathrm{MA}}$. Using Eq.~\eqref{eq:rhoMDW}, this equation is further written as $\partial_t(\rho_\mathrm{a0}c^2+W_\mathrm{int})=\phi_{\mathrm{MA}}$. The time derivative of the equilibrium mass density $\rho_\mathrm{a0}$ is zero, so we have $\partial_t W_\mathrm{int}=\phi_{\mathrm{MA}}$, which is equivalent to the definition of the power density in Eq.~\eqref{eq:phi_ma}. Thus, we see that, regarding the energy continuity equation, the mathematical difference between the MP and MMP formulations lies in how the interaction energy is treated -- explicitly in terms of the fields or implicitly as a part of the time-dependent mass density of the material, respectively.


\subsection{Momentum continuity equation}
Keeping the same considerations and approximations employed in the energy continuity equation, the momentum continuity equation, on its turn, is given in the laboratory frame as
\begin{eqnarray}\label{eq:f_MMP}  
\sum_{i=1}^{3}\hat{\bb e}_i\partial_{\nu}\mathcal{T}_{\mathrm{tot}}^{i\nu}=-\bb f_{\mathrm{MA}}-\bb f_{\mathrm{b}}  +\overleftrightarrow{\bm\nabla}\cdot(\rho_{\mathrm{a}}\bb v_{\mathrm{a}}\otimes \bb v_{\mathrm{a}})\nonumber \\
   +\frac{\partial}{\partial t}(\rho_{\mathrm{a}}\bb v_{\mathrm{a}})=0,
\end{eqnarray}
where $i=1,2,3$, $\hat{\bb e}_i$ is the unit vector in $i$-th direction and $\bb f_{\mathrm{b}}$ denotes the internal body force density in the medium, given by the spatial derivative of $\mathcal T_{\mathrm{mech}}^{i\nu}$. In the earlier example of a perfect fluid, we have $\bb f_{\mathrm{b}}=-\bm\nabla\mathcal P$. Besides this term, the additional difference in Eq.~(\ref{eq:f_MMP}) compared to the original MP formulation is in the electro- and magnetostriction forces contained in $\bb f_{\mathrm{MA}}$. Notice that, by using the material derivative $\dd/\dd t = \partial_t+\bb v_\mathrm a \cdot \bm \nabla$ and the continuity equation of the atomic number density $n_\mathrm{a}$, given by $\partial_tn_\mathrm{a}+\bm\nabla\cdot(n_\mathrm{a}\bb v_\mathrm{a})=0$, Eq.~(\ref{eq:f_MMP}) reduces to Newton's law of motion as $n_\mathrm a\dd \bb p_\mathrm a/ \dd t = \bb f_{\mathrm {MA}}+\bb f_\mathrm b$, where $\bb p_\mathrm{a}=\rho_\mathrm{a}\bb v_\mathrm{a}/n_\mathrm{a}$ is the momentum of a single atom}.

At last, the coupling between field and matter expected to arise in experimental observations, as already discussed for the energy case, is more subtle for momentum and will be addressed in the next section.


\section{Discussion}\label{sec:disc}

The general validity of Maxwell's SEM tensor as the electromagnetic part in coupled field plus matter systems with arbitrary sources, suggested in Sec.~\ref{sec:MA_SET}, 
is supported by the literature in works addressing electromagnetic force and torque with diverse source models, such as dipoles~\cite{Shen2023,Chaumet2000}, multipoles~\cite{Strasser2022,Wei2022}, negative-index scatterers~\cite{Chaumet2009} and nanoparticles placed in metamaterials~\cite{Shalin2015} and plasmonic traps~\cite{Zaman2019,Ren2021}. In these works, the total electromagnetic force is calculated as appropriate surface integrals of Maxwell's SEM tensor as obtained by integrating the general relation $\partial_{\nu}\mathcal{T}_{\mathrm{Maxwell}}^{\mu\nu} = -f^{\mu}$. However, some works employ Minkowski's SEM tensor (which generates the same force density as Abraham's when time-averaged) -- see, for example, Refs.~\cite{Gao2017,Li2020,Ullery2018}. These works, on their turn, are also subject to the problem discussed in Sec.~\ref{sec:MA_SET} of implicitly obtaining force densities that are valid only for macroscopic, effective medium descriptions -- nevertheless, for applications where only the system center of mass-energy movement is important and the total force is observed as a cycle-averaged quantity, such as in typical optical traps, this approach is consistent as well and agrees with our proposal~\cite{Kemp2017,Kemp2011}. 

As stated in Sec.~\ref{sec:MP+MA0}, the self-field interactions and their effects are excluded in the presented theory. 
To justify this, we refer to Ref.~\cite{Stabler1964}, where it was shown an alternative version of the SEM tensor for classical electromagnetic sources, which explicitly excludes self-field contributions. The equation of motion obtained from this modified tensor description is exactly the one predicted by the Lorentz force law but without the self-field contributions. Therefore, as we are not interested in radiation reaction phenomena, the Lorentz force law can be applied with no further complications, keeping in mind that in our system, the motion of a single dipole is always caused by the fields originating from all the other external sources.

The essential coupling between field and matter in the context of the Abraham-Minkowski problem has been discussed extensively in the literature, where different subsystem's separations (even arbitrary ones) were proposed~\cite{Barnett2010,Penfield1967,Kemp2011,Milonni2010}. In Sec.~\ref{sec:en}, the coupling in energy for the MMP formulation was discussed in terms of the fields and interaction energy densities. A similar but more subtle coupling is also present in the momentum continuity equation, Eq.~(\ref{eq:f_MMP}). By using the MP quasi-particle, it can be shown that in the laboratory frame the coupling between field and matter leads to a momentum transfer proportional to $n$~\cite{MP1,MP4} -- i.e., a Minkowski-type momentum, in accordance with the few existing measurements of photon momentum in matter~\cite{Jones1954,Jones1978,Strait2019,Campbell2005,Gibson1980}. This result is unambiguously tied to the field momentum carrying a momentum of Abraham type, proportional to $1/n$, while the matter contribution is attributed to the MDW carrying the difference, proportional to $(n-1/n)$. Therefore, as in the case of energy, our proposed theory agrees with the experimental works observing the total, coupled linear momentum of the field plus matter system. 
Indeed, as experimental observations in this scenario will always capture the closed system, without further argumentation, it is reasonable to assume that the decomposition of the total SEM tensor is arbitrary, as done for example in Ref.~\cite{Kemp2011}; nevertheless, the decompositions presented by us in Eqs.~(\ref{eq:T_MMP}) and (\ref{eq:T_tot}) have the advantage of describing the subsystems in terms of very well-known tensors, providing, therefore, a clearer physical interpretation. Lastly, recall that although $\mathcal{T}_{\mathrm{Maxwell}}$, $\mathcal{T}_{\mathrm{MDW}}$ and $\mathcal{T}_{\mathrm{mech}}$ are true tensors, the electromagnetic sources and consequently the optical force and power densities, given in Eqs.~(\ref{eq:f3}), (\ref{eq:4J}) and (\ref{eq:phi_ma}) respectively, were given specifically for the laboratory frame and non-relativistic velocities. Therefore, the presented theory is not covariant, but this generalization is of technical interest and will be addressed in future works.

From the definition of potential energy, the stress acting inside the material due to the electromagnetic interaction with the dipoles can be found from the derivative of the total (fields plus interaction) energy density, $(\bb D\cdot \bb E + \bb B\cdot\bb H)/2$, with respect to the associated strain component (in solid media)~\cite{Kittel}. 
It can be shown~\cite{Partanen2023} that 
such calculation leads to the force density $\bm \nabla (\bb P\cdot \bb E + \bb M \cdot \bb B)/2$, i.e., the work done by the electromagnetic field generates strain inside the material, which is manifested through the electro- and magnetostriction effects. The same occurs for fluid media, with strain being substituted by a local pressure variation. 
We can then see that striction effects correspond to pure stresses, and so they do not change the local momentum transfer from light inside the material, which occurs exclusively in the pulse propagation direction. Indeed, striction effects are long-known to not contribute to the material's center of mass-energy movement, 
having significant effects only in local force considerations~\cite{Brevik1979}. This interpretation also matches the discrete picture of light, where the incident photons on an isotropic dielectric have their average linear momentum in the light's propagation direction.

Eq.~(\ref{eq:f_MMP}) is compatible with the opto-elastic simulations from Refs.~\cite{MP1,Partanen2022b} -- striction effects are balanced by the strain-induced elastic forces, while the remaining force density is equal to $\bb f_{\mathrm{Ab}}$. This force density is responsible for the MDW propagating along with the light pulse. These dynamics and the linear momentum conservation can be visualized in detail in the two-dimensional simulations presented for an elastic solid medium in Ref.~\cite{Partanen2022b}. Another relevant fact is that an acoustic pressure wave would develop in fluids under pulsed optical excitation to counterbalance the effects of the generated striction force densities. This mechanism of acoustic relaxation was studied experimentally for non-magnetic fluids in Refs.~\cite{Astrath2022,Astrath2023,Astrath2024}, providing observations of the spatio-temporal dependence of the optical force density. We recall that, among all the known formulations for optical forces up to date, Eq.~(\ref{eq:f3}) is the only one capable of describing the experimental results from Refs.~\cite{Astrath2022,Astrath2023,Astrath2024} while being theoretically consistent with the observed model for magnetization in matter~\cite{Jackson1977,B,Hughes1951,Anghinoni2023}, as stated in Sec.~\ref{sec:intro}. This fact together with the familiar energy continuity equations shown in Eqs.~(\ref{eq:MPMA_en}) and~(\ref{eq:en2}) provide very good evidence for the validity of the presented theory. For a detailed discussion on the different existing formulations for electromagnetic energy and momentum transfer in matter, see for example Refs.~\cite{Brevik1979,Kemp2011,Anghinoni2022,Penfield1967}.

Although, as just discussed, good agreement with existing experiments has been found, the complete experimental confirmation of the presented theory requires the measurement of the small mass transferred by the mass density wave~\cite{MP1}. Such mass, denoted by $\delta m$, is given as $\delta m=\int \rho_{\mathrm{MDW}}(\bb r,t) \dd^3\bb r$, where the integration region covers the whole material. While optimized experimental setups to measure this quantity have been suggested~\cite{Partanen2022c,Partanen2019}, this measurement is quite challenging and has not, to our knowledge, been performed yet. In the MMP theory, $\rho_{\mathrm{MDW}}$ is given in Eq.~(\ref{eq:rhoMDW}) in terms of two contributions: one related to the small atomic displacements and the other one related to the mass equivalent of the energy spent in forming the induced dipoles configuration. The latter contribution does not generate a mass transfer -- therefore, $\delta m$ in the present theory is determined solely in terms of the atomic displacements. As shown in Ref.~\cite{Anghinoni2022}, $\bm \nabla\cdot \bb r_\mathrm a$ can be related to the field energy density distribution, yielding in the current case $\rho_{\mathrm{MDW}} \approx c^{-2}(n^2-1)\int_{-\infty}^t(\partial_{t'} W_{\mathrm{Maxwell}}+\phi_\mathrm{MA})dt'=c^{-2}(n^2-1)W_{\mathrm{Ab}}$. This is the exact same result as the original MP formulation~\cite{MP1,Anghinoni2022} -- therefore, $\delta m$ is unchanged in the MMP theory.

A covariant theory of light propagating in dielectric media using the MP formulation has already been reported~\cite{MP6,Partanen2023}. Specifically, in Ref.~\cite{Partanen2023}, a decomposition of the total SEM tensor very similar to the one shown in Eq.~(\ref{eq:T_MMP}) was adopted -- however, the field part was given by Abraham's SEM tensor, while this work adopts Maxwell's SEM tensor. In truth, whenever effective macroscopic treatments for the matter part are appropriate $\bb f_{\mathrm{Ab}}$ does originate from Abraham's SEM tensor (as discussed in Sec.~\ref{sec:MP+MA2}) -- in these cases, the MP formulation with the addition of striction effects, as described in Ref.~\cite{Partanen2023}, yields the exact same results as the MMP formulation. This occurs because the total force acting on the medium is the same in both MP and MMP formulations, i.e., $\int \bb f_{\mathrm{Ab}} \mathrm{d^3}\bb r=\int \bb f_{\mathrm{MA}} \mathrm{d^3}\bb r$ since striction forces always integrate to zero~\cite{Brevik1979,Kemp2017}. Therefore, the MMP formulation shares all the momentum and mass transfer properties and center of mass-energy movement of the MP formulation, which have been extensively discussed and shown to be mathematically consistent~\cite{MP1,MP2,MP3,MP4,MP5,MP6}. As a consequence, the MP and MMP formulations are not mutually exclusive: all the force densities and simulations from Refs.~\cite{MP1,Partanen2023} are correct, but the direct association of Abraham's SEM to $\bb f_{\mathrm{Ab}}$ is implicitly valid only for the conditions just cited. In this context, the MMP formulation can be seen as a clarification of the MP formulation, where the specific microscopic mechanisms of optical energy and momentum transfer involved are described in terms of electric and magnetic dipoles, eliminating the necessity of phenomenological approaches.

Having treated energy and linear momentum conservation, it is natural to ponder whether the presented theory is compatible with angular momentum conservation. Indeed, this is a very interesting topic that has drawn much attention in the past few decades, especially because it is now known that structured fields can carry both orbital and spin angular momentum~\cite{Allen1992,Barnett1994,Barnett2017,Bliokh2015,Yang2021,Nieto-Vesperinas2015}. 
In this scenario, simulations of optical angular momentum transfer using the MP formulation were already reported and seen to be in accordance with the conservation principle~\cite{MP3}. As striction forces are not related to linear momentum transfer, it is expected that this behavior is also shared by the MMP formulation.

The consideration of temporally dispersive media within the mass-polariton dynamics has been treated earlier~\cite{MP2}, where the momentum transfer was found to depend on both the wave's group and phase velocity. This is also expected to occur in the MMP formulation, but more careful analyses are needed because, for example, the power density shown in Eq.~(\ref{eq:phi_ma}) is not valid for dispersive media. Besides dispersion, the extension of the theory to include non-conservative optical forces~\cite{Sukhov2017}, anisotropy, and nonlinearities are also of interest. Additionally, a more fundamental formulation can be sought by working in a quantum mechanics regime, as done for example in Refs.~\cite{Loudon2002,LeFournis2022,Feigel2004}.

At last, recall that, as discussed, when electro- and magnetostriction force densities are counter-balanced by optically-induced stresses in the medium and the interaction energy density is accounted for explicitly, the equations for energy density and force density in the dielectric are equal to Abraham's formulation -- however, we emphasize that in local descriptions of energy and momentum, there is no physically meaningful association of these conserved quantities to Abraham's SEM tensor as it necessarily employs macroscopic sources in its application~\cite{Anghinoni2022,Milonni2010}. Therefore, the argumentation developed here reveals the true microscopic origin of Abraham's force density in linear, isotropic, lossless and non-dispersive dielectric media: it results from the optical force acting on moving induced dipoles partially counter-balanced by local, internal mechanical stresses. To the best of our knowledge, this simple interpretation of the long-known and widely employed Abraham force density in terms of microscopic dipoles had not been reported in the literature before. Clarifying this subtlety contributes to our fundamental understanding of the interaction of electromagnetic fields with matter and its numerous applications.

\section{Conclusions}\label{sec:conc}

We have described a microscopic
theory for the local electromagnetic energy and momentum transfer inside linear dielectric media called microscopic mass-polariton formulation. Our work unifies two recently introduced models of light in dielectric media: the microscopic Ampère formulation, based on classical dipolar electromagnetic sources, and the theoretical mass-polariton quasi-particle. This was made possible by noticing that the widely known Maxwell's stress-energy-momentum tensor is the only one capable of describing arbitrary electromagnetic sources in matter. The field part of the system is then given by this tensor, while the matter part is given by the mass density wave tensor, which corresponds to the disturbances in atomic positions propagating along with the electromagnetic wave, and a general mechanical tensor, which encompasses all effects not related to rest mass or kinetic energies, such as internal stresses. Additionally, in our model the electro- and magnetostriction effects are counter-balanced by the medium's local stresses, with Abraham's force density corresponding exactly to the remaining terms of the optical force density -- therefore, we have shown that whenever \textit{local} continuity considerations must take place Abraham's force density does not originate from Abraham's stress-energy-momentum tensor. This significantly improves our current knowledge of the field and matter interactions and also sheds light on the centenary Abraham-Minkowski controversy on the photon's momentum.

Our proposed theory
agrees with most experiments reported so far involving optical forces and radiation pressure in dielectric media, providing a paradigm for general opto-mechanical applications. When compared to the original mass-polariton formulation, our theory presents the advantage of being directly related to the very well-established dipolar model for electromagnetic sources in matter, which clarifies the microscopic dynamics of the system. Besides, electro- and magnetostriction effects are also naturally accounted for. 
The complete confirmation of our theory, however, still requires the measurement of the small mass transferred by the propagating mass density wave, which remains a very difficult experimental challenge and has not, to our knowledge, been performed yet.

\begin{acknowledgments}
The research leading to these results received funding from CNPq (304738/2019-0), CAPES (Finance Code 001), Funda\c{c}\~{a}o Arauc\'{a}ria, and FINEP. M.P. acknowledges funding from the Research Council of Finland under Contract No. 349971.
\end{acknowledgments}

\appendix

\section{Continuity equation for $\mathcal{T}_{\mathrm{Maxwell}}$}\label{sec:app_MA}

In covariant form, Maxwell's SEM tensor from Eq.~(\ref{eq:T_MA}) is given as~\cite{Luscombe,LL_fields}
\begin{align}
 \mathcal{T}_{\mathrm{Maxwell}}^{\mu\nu}
 =-\frac{1}{\mu_0}\left(\eta_{\alpha\lambda} F^{\mu\alpha}F^{\lambda\nu}+\frac{1}{4}\eta^{\mu\nu}F_{\alpha\beta}F^{\alpha\beta}\right).
\end{align}
Here, $\eta^{\mu\nu}$ is Minkowski's metric in $(-,+,+,+)$ signature and $F^{\mu\nu}$ is the usual electromagnetic field tensor, given in terms of the electromagnetic four-potential $A^{\mu}$ as $F^{\mu\nu}=\partial^{\mu} A^{\nu}-\partial^{\nu} A^{\mu}$. By taking the four-divergence of the above equation, we get~\cite{Luscombe,LL_fields}
\begin{equation}    \partial_{\nu}\mathcal{T}^{\mu\nu}_{\mathrm{Maxwell}}=-J_{\alpha}F^{\mu\alpha}.
\end{equation}
In terms of electromagnetic fields and sources, the right-hand side of the above equation is (apart from the minus sign) the
four-force $f^{\mu}=(\bb J\cdot\bb E/c,\rho \bb E+\bb J\times \bb B)$, from where we can recognize the Lorentz force density. By substituting into this equation the four-current from Eq.~(\ref{eq:4J}) and properly integrating over the volume, we obtain~\cite{Anghinoni2023} Eqs.~(\ref{eq:f3}) and~(\ref{eq:phi_ma}) for the force and power densities, respectively, in the MA formulation.

\section{Calculation of $\phi_{\mathrm{MA}}$}\label{app:phi}

The total power in the MA formulation is
\begin{equation}
    \Phi_{\mathrm{MA}} = \int \bb E \cdot \left[\dot{\bb p}\delta^3(\bb r) - \bb m \times \bm \nabla \delta^3(\bb r)\right] \dd ^3 \bb r.
\end{equation}

Adopting index notation for the last term, this equation is given by
\begin{equation}
    \Phi_{\mathrm{MA}} = \dot{\bb p} \cdot\bb E-\int E_i \epsilon_{ijk}m_j \partial_k \delta^3(\bb r) \dd^3 \bb r,
\end{equation}
where $\epsilon_{ijk}$ is the Levi-Civita symbol and the summations in repeated indices are implied. Integrating the last term by parts 
we obtain
\begin{equation}
    \Phi_{\mathrm{MA}} = \partial_t \bb p \cdot\bb E+\epsilon_{ijk} m_j \partial_k E_i,
\end{equation}
where we took $\dot{\bb p} \approx \partial_t \bb p$ as the dipole's velocity in the laboratory frame is much smaller than $c$. In vector notation, the second term on the right-hand side is equal to $\bb m \cdot \left(\bm \nabla \times \bb E\right)$. Invoking Faraday's law, we have $\Phi_{\mathrm{MA}} = \partial_t \bb p \cdot\bb E - \bb m \cdot \partial_t \bb B$. Notice, however, that this result for $\Phi_{\mathrm{MA}}$ must be valid for the laboratory inertial reference frame. In this frame, the dipole moments are given, to first order in velocities, as~\cite{Hnizdo2012} $\bb p = \bb p_0+\bb v \times \bb m_0/c^2$ and $\bb m = \bb m_0 -\bb v \times \bb p_0$, where $\bb p_0$ and $\bb m_0$ are the electric and magnetic dipole moments in the dipole's rest frame, respectively, and $\bb v$ is the dipole's velocity in the laboratory frame. On their turn, the fields are already assumed to be given in the laboratory frame. Therefore, we have
\begin{equation}
    \Phi_{\mathrm{MA}} = \partial_t (\bb p_0+\bb v \times \bb m_0/c^2) \cdot\bb E - (\bb m_0 -\bb v \times \bb p_0)\cdot \partial_t \bb B.
\end{equation}
This result represents the contribution to the total power from one single dipole moving with velocity $\bb v$ inside the integration volume. By summing the contribution from all dipoles in the integration volume and dropping the unnecessary subscripts, we obtain
\begin{equation}\label{eq:phi2}
    \phi_{\mathrm{MA}} = \partial_t (\bb P+\bb v\times \bb M/c^2) \cdot\bb E- (\bb M-\bb v\times \bb P) \cdot \partial_t \bb B,
\end{equation}
where $\bb P$ and $\bb M$ correspond, respectively, to $\bb p$ and $\bb m$ divided by the integration volume and multiplied by the number of dipoles in this volume. Naturally, the above equation is based on the traditional description adopted in condensed matter where macroscopic fields are built from their associated microscopic counterparts -- in our case, the polarization $\bb P$ and magnetization $\bb M$ and the dipole moments $\bb p$ and $\bb m$, respectively -- through a suitable spatial smoothing procedure.

Recalling that $|\bb v| \ll c$, we may neglect the terms proportional to $\bb v$ in Eq.~(\ref{eq:phi2}). Besides, as the medium is assumed linear and non-dispersive, we have at last
\begin{equation}
    \phi_{\mathrm{MA}} \approx \frac{1}{2}\frac{\partial}{\partial t}\left({\bb P} \cdot \bb E -\bb M \cdot  \bb B\right),
\end{equation}
which is exactly Eq.~(\ref{eq:phi_ma}).

\section{The MMP Lagrangian and hidden momentum contribution}\label{sec:HM}

To justify the inclusion of the hidden energy and momentum contributions in Eqs.~(\ref{eq:MPMA_en}) and (\ref{eq:f_MMP}), respectively, we will initially follow the main steps from the derivation of the Lagrangian given in Ref.~\cite{Horsley2006}. The starting point is the classical version of the well-known Quantum Electrodynamics (QED) Lagrangian for spinless, charged point particles interacting within external electromagnetic fields. As we are interested in electromagnetic dipolar sources, we will consider two point particles with identical mass $m$ and opposite elementary charges $e$. The QED Lagrangian in this case reads
\begin{eqnarray}\label{eq:L}
    L = -m c^2\left[\left(1-|\dot{\bb r}_1|^2/c^2\right)^{1/2}+\left(1-|\dot{\bb r}_2|^2/c^2\right)^{1/2}\right] \nonumber \\
    - \int \left[J_{\mu}A^{\mu}+\frac{\varepsilon_0}{4}F_{\mu\nu}F^{\mu\nu}\right]\mathrm{d}^3\bb r.
\end{eqnarray}
Here, $\bb r_\mathrm{i}$ and $\dot{\bb r}_\mathrm{i}$, $i=1,2$, are the position and velocity of the $i$-th particle, respectively, as given in the laboratory inertial reference frame, and $A^{\mu}=(\varphi/c,\bb A)$ is the electromagnetic four-potential. The electrical charge density is $\rho(\bb r)=e\delta^3(\bb r - \bb r_1)-e\delta^3(\bb r-\bb r_2)$ and the electrical current density is $\bb J(\bb r)=e\dot{\bb r}_1\delta^3(\bb r - \bb r_1)-e\dot{\bb r}_2\delta^3(\bb r-\bb r_2)$, which are equivalent to the four-current in Eq.~(\ref{eq:4J}) for large observation distances $|\bb r|$. 

Next, the Power-Zienau-Woolley (PZW) gauge transformation~\cite{Power1959,Woolley1971,Babiker1983} is applied to the Lagrangian in Eq.~(\ref{eq:L}), allowing us to write the modified Lagrangian $L'$ in terms of field quantities as
\begin{eqnarray}\label{eq:L2}
    L'&=& -m c^2\left[\left(1-|\dot{\bb r}_1|^2/c^2\right)^{1/2}+\left(1-|\dot{\bb r}_2|^2/c^2\right)^{1/2}\right] \nonumber \\
    &+&\int \left\{ \frac{\varepsilon_0}{2}\left[|\bb E(\bb r)|^2-c^2|\bb B(\bb r)|^2\right]+ \frac{1}{2}\bb P(\bb r)\cdot \bb E(\bb r)\nonumber \right.\\
    &&\left.+\frac{1}{2}\bb M(\bb r)\cdot \bb B(\bb r)- \dot{\bb R}\cdot\left[\bb P(\bb r)\times \bb B(\bb r)\right]\right\} \mathrm{d}^3 \bb r.
\end{eqnarray}
In the above equation, $\bb R=(\bb r_1+\bb r_2)/2$ is the center of mass-energy of the system as measured in the laboratory frame. The first two terms on the right-hand side are not affected by the PZW transformation. The first term inside the integral is an invariant of the electromagnetic field and is related to the field's energy density. The second and third ones are associated to the dipolar interaction and, consequently, to electro- and magnetostriction effects (actually $\bb P$ and $\bb M$ are initially related to the complete multipole expansions~\cite{Horsley2006}, but we have already truncated it to account only for dipolar interactions). The 1/2 factor occurs due to the linearly induced dipoles by the fields.
The last term is a linear momentum contribution known as R\"ontgen interaction and appears naturally even in non-relativistic derivations~\cite{Stenholm1986,Lembessis1993,Anghinoni2023}.

Now, we promote $\bb R$ to a dynamical variable of the system and introduce the relative coordinate $\bb q = \bb r_2 - \bb r_1$. To leading order in the velocities divided by $c$, it can be shown the Lagrangian in Eq.~(\ref{eq:L2}) becomes
\begin{eqnarray}\label{eq:L3}
    L'&=& -Mc^2+\frac{M|\dot{\bb R}|^2}{2}+\frac{m_{\mathrm{red}}|\dot{\bb q}|^2}{2} \nonumber \\
    &&+\int \biggr[ \frac{\varepsilon_0}{2}\left(|\bb E(\bb r)|^2-c^2|\bb B(\bb r)|^2\right)+ \frac{1}{2}\bb P(\bb r)\cdot \bb E(\bb r) \nonumber\\
    &&\left.+\frac{1}{2}\bb M(\bb r)\cdot \bb B(\bb r)- \dot{\bb R}\cdot\left(\bb P(\bb r)\times \bb B(\bb r)\right)\right.\nonumber\\
    &&+ \frac{\dot{\bb R}}{c^2}\cdot\left(\bb M(\bb r)\times \bb E(\bb r)\right)\biggr] \mathrm{d}^3 \bb r,
\end{eqnarray}
where $M = 2m$ is the total mass of the system and $m_{\mathrm{red}}=m/2$ is the reduced mass. The last term in the integral is directly related to the hidden momentum and hidden energy contributions (which are also known as Aharonov-Casher interaction in the QED context), and arises as a first order effect ($\gamma \approx 1$) of the relativistic Lagrangian.

In the MMP formulation, the dielectric medium is described under the continuum approximation as a distribution of point dipoles -- this way, we have $\bb q \to 0$, and the third term on the right-hand side of Eq.~(\ref{eq:L3}) is zero. By introducing the atomic mass density $\rho_\mathrm{a}$, the remaining mechanical terms are transformed as 
\begin{eqnarray}
  -M c^2  \to  -\int \rho_\mathrm{a}(\bb r) c^2\mathrm{d}^3\bb r
\end{eqnarray}
and
\begin{eqnarray}
\frac{M|\dot{\bb R}|^2}{2} \to \frac{1}{2}\int \rho_\mathrm{a}(\bb r) |\bb v_\mathrm{a}(\bb r)|^2\mathrm{d}^3\bb r,
\end{eqnarray}
which are associated to the rest and kinetic energies of the medium, respectively, with $\dot{\bb R}$ being identified as the atomic velocity $\bb v_{\mathrm{a}}$. With this continuum description and recalling the kinetic energy term is negligible, we can then finally write the Lagrangian density for the MMP formulation as
\begin{eqnarray}\label{eq:L4}
    \mathcal{L}_{\mathrm{MMP}} &=& -\rho_\mathrm{a} c^2 
    +\frac{1}{2}\bb D\cdot\bb E-\frac{1}{2}\bb B\cdot\bb H
    - \dot{\bb R}\!\cdot\!\left(\bb P\!\times\! \bb B\right)\nonumber\\
    & &+ \frac{\dot{\bb R}}{c^2}\cdot\left(\bb M\times \bb E\right),
\end{eqnarray}
which is valid for the laboratory frame and approximated to leading order in the velocities, with the dependence on $\bb r$ omitted for brevity. The MDW mass density is explicitly obtained in the analysis when we study how much the original mass density $\rho_\mathrm{a0}$ becomes disturbed by the optical force as $\rho_\mathrm{MDW}=\rho_\mathrm{a}-\rho_\mathrm{a0}$. 

The momentum density of the MDW can be obtained by varying the action integral in Eq.~(\ref{eq:L4}) with respect to $\dot{\mathbf{R}}$ as $\partial\mathcal{L}_\mathrm{MMP}/\partial\dot{\mathbf{R}}=-\mathbf{P}\times\mathbf{B}+\frac{1}{c^2}\mathbf{M}\times\mathbf{E}$, in which the hidden momentum contribution is transparently present. Alternatively, if the kinetic energy density term $\rho_\mathrm{a} |\bb v_\mathrm{a}|^2/2$ is kept, the equations of motion yield $\rho_\mathrm{a}\bb{v}_\mathrm{a}-\mathbf{P}\times\mathbf{B}+\frac{1}{c^2}\mathbf{M}\times\mathbf{E}=0$, which is an equation for the MDW momentum density as $\rho_\mathrm{MDW}\bb{v}_\mathrm{l}=\rho_\mathrm{a}\bb{v}_\mathrm{a}=\mathbf{P}\times\mathbf{B}-\frac{1}{c^2}\mathbf{M}\times\mathbf{E}$. 


\bibliography{Refs}

\end{document}